\documentclass[intlimits,twoside,a4paper]{article}

\usepackage{amsmath,amssymb}
\usepackage{graphicx}

\usepackage[T2A]{fontenc}
\usepackage[cp1251]{inputenc}

\usepackage[eqsecnum]{cmpj2}

\usepackage{makeidx}
\usepackage{bm}
\usepackage{amssymb}
\usepackage{amssymb,amsmath}
\usepackage{graphicx}
\usepackage{graphics}
\usepackage{dsfont}
\usepackage{slashed}
\usepackage{color}
\usepackage[normalem,normalbf]{ulem}
\usepackage{cancel}

\newcommand{\nn}{\nonumber}

\issue{2017}{20}{1}{13603}
\doinumber{10.5488/CMP.20.13603}

\title[Emergent universal critical  behavior]{Emergent universal critical  behavior of the 2D $N$-color Ashkin-Teller  model
in the presence of correlated disorder}
\author[M. Dudka, A.A. Fedorenko]{M. Dudka\refaddr{label1}, A.A. Fedorenko\refaddr{label2}}
\addresses{
\addr{label1} Institute for Condensed Matter Physics
of the National Academy of Sciences of Ukraine,\\
1 Svientsitskii St., 79011 Lviv, Ukraine
\addr{label2} Laboratoire de Physique, ENS de Lyon, Univ Claude Bernard, Univ Lyon, CNRS,  F-69342 Lyon, France}

\authorcopyright{M. Dudka, A.A. Fedorenko, 2017}
\date{Received January 12, 2017, in final form February 14, 2017}

\begin{document}

\maketitle

\begin{abstract}
We study the critical behavior of the 2D $N$-color Ashkin-Teller  model
in the presence of random bond disorder whose correlations decays
with the distance $r$ as a power-law  $r^{-a}$. We consider the case when the spins
of different colors sitting at the same site are coupled
by the same bond  and map this problem onto the 2D system of $N/2$ flavors of
interacting Dirac fermions in the presence of correlated disorder.
Using renormalization group we show that for $N=2$,
a ``weakly universal'' scaling behavior at the
continuous transition becomes  universal with  new critical exponents.
For $N>2$,  the first-order phase transition is rounded by the correlated
disorder and turns into a continuous one.
\keywords phase transitions, correlated disorder, two-dimensional models, Dirac fermions, renormalization group
\pacs 61.43.-j, 64.60.ae, 64.60.Bd, 75.10.Hk

\end{abstract}

\section{Introduction}

Two-dimensional (2D) systems are of particular interest in
studying  phase transitions in condensed matter.
On the one hand, this interest is constantly growing due to the progress in
experimental techniques of producing and studying  low-dimensional
materials like graphene~\cite{Novoselov:05},
2D crystals \cite{2Dcrystals}, and
ultrathin ferromagnetic films \cite{films}.
On the other hand,  the scaling  behavior of
many models is much easier
to analyze in two dimensions than in three dimensions,
since the 2D conformal invariance much stronger restricts possible scenarios,
at least in the absence of  disorder \cite{Bernard:1995}.
Moreover,
some of 2D models allow for an exact solution. The well-known examples include
the 2D Ising model~\cite{Onsager},
the 2D ice-model (6-vertex model)~\cite{Fan:1970} and
the Baxter  model (symmetric 8-vertex model)~\cite{Baxter}.
The latter model is related to the so-called  Ashkin-Teller
model~\cite{AshkinTeller}, which was introduced to describe cooperative phenomena in
quaternary alloys~\cite{Wegner:1972}, along a selfdual line.
Both models can be described in terms of two Ising models coupled
through their energy densities (i.e., by four-spin interaction):
\begin{eqnarray} \label{ash-tel1}
H =  -\sum_{\langle r,r'\rangle} \left[ J\left(\sigma_r^{1}\sigma_{r'}^{1}+\sigma_r^{2}\sigma_{r'}^{2}\right)+J_4\sigma^1_r\sigma^1_{r'}\sigma^2_r\sigma^2_{r'}\right].
\end{eqnarray}
The main feature of the model (\ref{ash-tel1}) is a continuous transition
with the so-called ``weak universality'' \cite{Suzuki}: contrary to the usual
critical universality, the critical exponents of model (\ref{ash-tel1})
continuously depend on  the coupling constant $J_4$.
In particular, the correlation length exponent is \cite{Baxter}
\begin{equation}\label{nup}
\frac1{\nu_{\text{pure}}} = \frac4{\pi} \arctan\left(\re^{2 J_4/T_{\text c}}\right),
\end{equation}
where the critical temperature  of the pure system $T_{\text c}$ is given by $2J/T_{\text c} = \ln(1+\sqrt{2})$.
The heat capacity exponent behaves as $\alpha_{\text{pure}} \sim J_4$, and thus, changes
sign with $J_4$.
However, expressing  the singular behavior of thermodynamic quantities
near the critical point in terms of the inverse correlation length
rather than of the reduced temperature, one finds that
the  critical exponents rescaled in this way are universal~\cite{Krcmar:2016}.

Generalizing the model (\ref{ash-tel1}) to an arbitrary $N$,
one arrives at  the  so-called $N$-color Ashkin-Teller  model~\cite{GrestWidom}:
\begin{equation}
\label{nash-tel}
H=-\sum_{\langle r  ,r' \rangle }\left( J \sum^N_{a=1}\sigma^a_r\sigma^a_{r'}
+J_4\sum^N_{a<b}\sigma^a_r
 \sigma^a_{r'}\sigma^b_r\sigma^b_{r'}\right),
\end{equation}
in which $N$ Ising models  are coupled pairwise through
interaction $J_4$. It reduces to the usual Ashkin-Teller (Baxter) model for $N=2$.
For $N>2$,  the transition properties of model (\ref{nash-tel})
drastically depend on the sign of $J_4$:
the system undergoes a continuous phase transition
for $J_4<0$,
while for $J_4>0$, the transition is of first-order~\cite{GrestWidom}.

The effects of quenched disorder on phase transitions
have been a hot topic of research for several decades
(for review see e.g., \cite{Pelissetto_rev,Dotsenko_rev, Holovatch_rev}).
For example, it is well-known that
the critical exponents of a system undergoing a continuous
phase transition may be modified by uncorrelated
quenched impurities coupled to the local
energy density (the so-called random-bond or random site disorder).
In this case,  the relevance of disorder can be predicted  using the Harris
criterion~\cite{Harris74}: if the heat capacity exponent
of the corresponding pure system is positive,
$\alpha_{\text{pure}}=2-d\nu_{\text{pure}}>0$, then
the presence of weak uncorrelated disorder leads to a new critical behavior.
Here, $\nu_{\text{pure}}$ is the correlation length exponent of the pure system.
According to the Harris criterion, the 2D Ising model corresponds to
a marginal situation  since the heat capacity exhibits only a logarithmic
divergence in the vicinity of critical point, i.e.,
$\alpha_{\text{pure}}=0$~\cite{Onsager}.
Explicit calculations performed for the disordered 2D Ising model
in \cite{DDf1,DDf2,DDf3,2DRIM1,2DRIM2} reveal
that uncorrelated  disorder modifies the logarithmic divergence to a
double logarithmic behavior, while other power-law scaling laws
acquire  universal logarithmic corrections. The situation is quite
different when the quenched
disorder is correlated.   According to the generalized
Harris criterion~\cite{weinrib-83},
the Gaussian disorder, whose variance decays as a power law $r^{-a}$,
modifies the critical behavior
of the pure system for $a < d$ provided that it satisfies
the inequality $\nu_{\text{pure}} < 2/a$.
For $a > d$, the usual Harris criterion
is recovered and the condition is replaced by
$\nu_{\text{pure}} < 2/d$ (see also \cite{HonkonenNalimov,Prudnikov:2000}).
For the 2D Ising model with long-range correlated disorder,
this has been explicitly shown in \cite{correlated}
by mapping to the 2D Dirac fermions.

The effect of uncorrelated disorder on the continuous phase
transition with ``weak universality'' exhibited by the model~(\ref{ash-tel1})
has been considered
in \cite{DD1,MatthewsMorgan1984,WisemanDomany19951,WisemanDomany19952,Shalaev:1994}.
Since the heat capacity exponent $\alpha_{\text{pure}}$  of the pure Baxter
model is positive for $J_4>0$, one can expect that the critical behavior
is modified  by uncorrelated disorder for $J_4>0$.
The  renormalization group (RG) picture obtained using a mapping to fermions
suggests~\cite{DD1,Shalaev:1994}
that for $J_4>0$, the ``weakly universal'' critical behavior of the Baxter model
changes to that of the disordered 2D Ising model, e.g.,
the heat capacity exhibits a double logarithmic singularity. The numerical
simulations of \cite{MatthewsMorgan1984,WisemanDomany19951,WisemanDomany19952}
 support the relevance of disorder but quantitatively they are
less conclusive. For instance,
the numerical simulations of the random bond Ashkin-Teller model
seem to require  additional efforts  due to large sample to sample
fluctuations
and are rather in favor of a
logarithmic than a double logarithmic behavior of the
heat capacity~\cite{WisemanDomany19951,WisemanDomany19952}.
For $J_4<0$, the exponent $\alpha_{\text{pure}}$ is also negative so that
the Harris criterion naively suggests that the critical behavior is unaffected
by uncorrelated disorder. The critical behavior deduced from the RG picture
is, however, different due to a new vertex generated by the RG flow.
For instance, the correlation length behavior becomes universal
while the heat capacity remains finite.
However, the precise behavior depends
on the initial disorder, in particular, whether  the disorder seen by the both
coupled Ising models  in equation~(\ref{ash-tel1}) is correlated or
not~\cite{DD1,Shalaev:1994}.

It seems rather striking that adding a weak short-range correlated (SR) or
uncorrelated quenched  disorder
to the 2D $N$-color Ahkin-Teller (\ref{nash-tel}) with $N>2$
results in emergent critical behavior~\cite{Murthy1987,Cardy1,Cardy2,first-to-second1,first-to-second2,first-to-second3,first-to-second4}.
Indeed, the pure model exhibits a fluctuation-driven first-order transition
characterized by runaway of the  RG flow which is reversed
by even weak uncorrelated  disorder. The 2D
three-color Ashkin-Teller model has been studied
by means of large-scale Monte Carlo simulations in \cite{Bellafard2012}
and~\cite{Zhu2015}. While the first early  work excludes
the possibility of continuous transition with universal scaling behavior,
the second recent paper demonstrates that the first-order
phase transition is rounded by the disorder and turns into a continuous one.
The resulting transition seems to be in the disordered  2D Ising universality class.
This agrees with perturbative RG predictions
of \cite{Murthy1987,Cardy1,Cardy2,first-to-second1,first-to-second2,first-to-second3,first-to-second4}.

 In the present paper we study the effects of long-range  correlated (LR)
disorder with power-law decay of correlations
on the phase transitions in the Baxter and $N$-color Ashkin-Teler models.
The article is organized as follows.
In  section~\ref{sec:models}, we  consider the formulation of the problem in terms of
Dirac (complex) fermions
following \cite{correlated,fedorenko2012,DD2} and restrict our
consideration to the case of the same disorder for
all fermion flavors [i.e., the disorder potentials seen by different
Ising components in the models (\ref{ash-tel1}) and (\ref{nash-tel})
are completely correlated].
In section~\ref{sec:models},  we introduce a fermion representation
for both models. In section~\ref{sec:RG} we briefly describe
the renormalization scheme we use for both models
while in section~\ref{sec:one-loop}
we present the one-loop RG functions and the derived scaling behavior.
Finally, we conclude in section~\ref{sec:conclusion}.

\section{Models and their fermion representation} \label{sec:models}

In the vicinity of a critical point, the long-distance properties
of the Baxter model (\ref{ash-tel1}) can be described
using two  Majorana (real) fermionic fields  $\chi_1$ and $\chi_2$ with the
action~\cite{DD1}:
\begin{eqnarray} \label{act_major}
S =
\int \rd^2 r \left\{ \frac{1}{2}\bar{\chi}_1[\slashed{\partial}  + m(r)  ]
  \chi_1 +\frac{1}{2}\bar{\chi}_2[\slashed{\partial}  + m(r)  ]
  \chi_2
- \frac{1}{4} \lambda_0
(\bar{\chi}_1\chi_1)
(\bar{\chi}_2 \chi_2)\right\}.\ \ \
\end{eqnarray}
Here, we define
$\slashed{\partial}=\gamma_1\partial_1+\gamma_2\partial_2$,
$\gamma_1=\sigma^x$, $\gamma_2=\sigma^y$, and $\bar{\chi}_i=\chi_i^T \sigma^y$
with $\sigma^x$, $\sigma^y$, $\sigma^z$ being the
Pauli matrices. In the clean case, we have $m(r)=m_0\sim(T_{\text c}-T)/T_{\text c}$ and $\lambda_0\sim J_4$.
The same model can be equivalently expressed  in terms of a single
Dirac (complex) fermionic field $\psi=(\chi_1+\ri\chi_2)/\sqrt{2}$ with
$\bar{\psi}=(\bar{\chi}_1-\ri\bar{\chi}_2)/\sqrt{2}$. The corresponding action reads~\cite{DD2}
\begin{eqnarray} \label{act_dirac}
S =
\int \rd^2 r \left\{ \bar{\psi}[\slashed{\partial}  + m(r)  ]
  \psi
- \frac{1}{2} \lambda_0
(\bar{\psi}\psi)
(\bar{\psi} \psi)\right\}.\ \ \
\end{eqnarray}
Generalization of the action (\ref{act_dirac}) to  the
$N$-color Ashkin-Teler model with \textit{even} $N$ is straightforward~\cite{Ludwig:1987}:
\begin{eqnarray} \label{act_diracN}
S =
\int \rd^2 r \left\{ \sum_{i=1}^{N/2}\bar{\psi}_i[\slashed{\partial}  + m(r)  ]
  \psi_i
- \frac{1}{2} \lambda_0\sum_{i,j=1}^{N/2}
(\bar{\psi}_i\psi_i)
(\bar{\psi}_j \psi_j)\right\},\ \ \
\end{eqnarray}
where we have introduced $N/2$ flavors of Dirac fermions instead of
$N$ flavors of Majorana fermions.

In the presence of random bond disorder which is completely correlated between
different Ising colors (i.e., different fermion flavors),  the mass can be written as
$m(r) = m_0 +\delta m(r)$, where $\delta m(r)$ is the local
disorder strength. We assume that the disorder strength
is a random Gaussian with the
mean value  $\overline{\delta m(r)}=0$ and variance
\begin{eqnarray} \label{eq:dis-cor-0}
\overline{\delta m(r)\delta m(0) }= g(r),
\end{eqnarray}
where $g(r)\sim \delta(r)$ and $g(r)\sim r^{-a}$ are for SR
and  LR  disorder, respectively.
We shall use dimensional regularization
so that we have to generalize the problem
to arbitrary $d$. To that end, we replace the Pauli matrices by the Clifford
algebra represented by the matrices $\gamma_i$, $i=1,\ldots,d$ satisfying the
corresponding anticommutation relations.
To average the free energy of (\ref{act_diracN}) over different disorder
configurations, we employ the replica trick~\cite{edwards_replica}. Introducing
$n$ replicas of the system (\ref{act_diracN}) and averaging over Gaussian
disorder distribution  we arrive at the replicated effective action
\begin{align} \label{act_diracNn}
S_{\rm eff} &=
\sum_{\alpha=1}^{n} \int \rd^d r \left[ \sum_{i=1}^{N/2}\bar{\psi}^\alpha_i(\slashed{\partial}  + m_0  )
  \psi^\alpha_i
- \frac{1}{2} \lambda_0 \sum_{i,j=1}^{N/2}
\left(\bar{\psi}^\alpha_i\psi^\alpha_i\right)
\left(\bar{\psi}^\alpha_j \psi^\alpha_j\right)\right]\nonumber\\
&\quad-\frac12 \sum_{\alpha,\beta=1}^{n}\int \rd^d r \int \rd^d r'
\sum_{i,j=1}^{N/2} g(r-r')\left[\bar{\psi}^\alpha_i(r)\psi^\alpha_i(r)\right]\left[\bar{\psi}^\beta_j(r')\psi^\beta_j(r')\right].
\end{align}
The properties of the
original system with quenched disorder can be then obtained
by taking the limit $n\to 0$. It is convenient to fix
the normalization of the disorder correlator~(\ref{eq:dis-cor-0})
in Fourier space as
\begin{eqnarray}
\overline{\delta m(k)\delta m(k') }= (2\pi)^d \delta^d(k+k') \bar{g}(k),
\end{eqnarray}
with
\begin{eqnarray}\label{furier}
\bar{g}(k) = u_{0} + v_{0} k^{a-d},
\end{eqnarray}
where $u_0$ and $v_0$ are bare coupling constants corresponding to the SR
and LR parts of disorder correlator,  respectively.

\section{RG description} \label{sec:RG}

We study the long-distance properties of (\ref{act_diracNn})  within the standard approach of field-theoretical  RG technique \cite{RG}.
Applying it  one can calculate the correlation
functions for the  action (\ref{act_diracNn})  perturbatively in $\lambda_0$, $u_{0}$ and  $v_{0}$.
The integrals entering this perturbation series turn out to be UV divergent in the dimension we are interested in ($d=2$).
To make the theory finite we use the dimensional regularization~\cite{HooftVelman1,HooftVelman2} and compute all integrals in $d=2-\varepsilon$.
Inspired by the works \cite{weinrib-83,fedorenko2012} we perform a double expansion in $\varepsilon=2-d$  and $\delta=2-a$ so that all divergences
are transformed into poles in $\varepsilon$  and $\delta$ while the ratio $\varepsilon/\delta$ remains finite.
We define the renormalized fields $\psi_i$, $\bar{\psi}_i$, mass $m$, and dimensionless coupling constants $\lambda$, $u$ and $v$
in  such a way that all poles can be hidden in the renormalization factors $Z_\psi$, $Z_m$,  $Z_\lambda$, $Z_{u}$ and $Z_{v}$ leaving finite
the correlation functions computed with the renormalized action
\begin{align} \label{eq:action-r}
  S_{\text R}\!\! &= \!\!\sum\limits_{\alpha{=}1}^n\sum\limits_{i{=}1}^{N/2}
\int_k \bar{\psi}^{\alpha}_i({-}k)(  Z_\psi \bm{\gamma} k  {-} Z_m  m  )
  \psi^{\alpha}_i(k) \nonumber\\
& \quad- \frac{1}{2}\!\!\sum\limits_{\alpha{=}1}^n\sum\limits_{i,j{=}1}^{N/2}\int_{k_1,k_2,k_3}\!\! \mu^{\varepsilon} Z_{\lambda} \lambda \bar{\psi}^{\alpha}_i(k_1) \psi^{\alpha}_i(k_2)
\bar{\psi}^{\alpha}_j(k_3) \psi^{\alpha}_j({-}k_1{-}k_2{-}k_3) \nonumber \\
&\quad
- \frac12\!\! \!\sum\limits_{\alpha,\beta=1}^n\! \sum\limits_{i,j{=}1}^{N/2}\int_{k_1,k_2,k_3} \!\!
\left( \mu^{\varepsilon} Z_{u} u + \mu^{\delta}Z_{v} v |k_1{+}k_2|^{a{-}d} \right)  \bar{\psi}^{\alpha}_i(k_1) \psi^{\alpha}_i(k_2)
\bar{\psi}^{\beta}_j(k_3) \psi^{\beta}_j({-}k_1{-}k_2{-}k_3),
\end{align}
where we have introduced a renormalization scale $\mu$ and the shortcut notation $\int_k  := \int \rd^d k$, so that $\int_{k_1,k_2,k_3}$ stands for the corresponding triple integral.
Since the renormalized action is obtained from the bare one by the fields rescaling
\begin{eqnarray}
\psi_{i,0} = Z_{\psi}^{1/2} \psi_{i}, \qquad  \bar{\psi}_{i,0} = Z_{\psi}^{1/2} \bar{\psi}_{i}\,,
\end{eqnarray}
the bare and renormalized parameters are related by
\begin{eqnarray}
m_0 = {Z_m}Z_{\psi}^{-1} m,  \qquad
\lambda_0 = \mu^{\varepsilon }  Z_{\lambda} Z_{\psi}^{-2} \lambda,  \\
u_{0} =  \mu^{\varepsilon }  Z_{u} Z_{\psi}^{-2} u,\qquad
v_{0} = \mu^{\delta }  Z_{v} Z_{\psi}^{-2} v,
\end{eqnarray}
where we have included $K_d/2$ in redefinition of $\lambda$, $u$ and $v$.
$K_d = 2\pi^{d/2}/(2\pi)^d\Gamma(d/2)$ is the area of the $d$-dimensional unit sphere divided by $(2\pi)^d$.
The renormalized $\mathcal N$-point vertex function ${\Gamma}^{(\mathcal N)}$  is related to
the bare ${\Gamma}_0^{(\mathcal N)}$ by
\begin{equation}
{\Gamma}_0^{(\mathcal N)}(k_i; m_0,u_0,v_0) = Z_{\psi}^{-{\mathcal N}/2}{\Gamma}^{(\mathcal N)}(k_i;  m ,u,v, \mu). \ \
\end{equation}
To calculate the renormalization constants it suffices to renormalize the two-point vertex function $\Gamma^{(2)}$ and
the four-point vertex function $\Gamma^{(4)}$. We impose that they are finite at $m=\mu$ and find the renormalization constants
using a minimal subtraction scheme~\cite{HooftVelman1,HooftVelman2}.  To that end, it is convenient to
split the four-point function in the clean  $\Gamma_\lambda$,  SR disorder $\Gamma_u$
and LR disorder $\Gamma_v$ parts:
\begin{eqnarray}
{\Gamma^{(4)}}^{ijkl}_{\alpha\beta\gamma\mu}(k_1,k_2,k_3,k_4){=}\left\{\Gamma_\lambda^{(4)}(k_i)\delta_{\alpha\beta}\delta_{\alpha\gamma}\delta_{\alpha\mu}{+}
\left[\Gamma_u^{(4)}(k_i){+}\Gamma_v^{(4)}(k_i)|k_1{+}k_2|^{a-d}\right]\delta_{\alpha\beta}\delta_{\gamma\mu}\right\}\delta_{ij}\delta_{kl}\,.
\end{eqnarray}
Renormalization constants are determined from the condition that
$\Gamma_\lambda^{(4)}(0;m=\mu)$, $\Gamma_u^{(4)}(0;m=\mu)$ and $\Gamma_v^{(4)}(0;m=\mu)$  are finite.

Since the bare vertex function does not depend on the renormalization scale $\mu$,
the renormalized vertex function satisfies the RG equation
\begin{eqnarray}
&&\left[\mu\frac{\partial}{\partial \mu}-\beta_{\lambda}(\lambda,u,v)
\frac{\partial}{\partial \lambda} -\beta_{u}(\lambda,u,v)
\frac{\partial}{\partial u} -  \beta_{v}(\lambda,u,v)
\frac{\partial}{\partial v} - \frac{\mathcal N}2 \eta_\psi(\lambda,u,v)  \right. \nn \\
&& \ \ \left.
 + \gamma(\lambda,u,v)  m \frac{\partial}{\partial m}\right]
  {\Gamma}^{(\mathcal N)}(k_i; m ,\lambda,u,v, \mu)=0,  \label{eq:RG1}
\end{eqnarray}
where we have introduced the RG functions
\begin{eqnarray}
&&\beta_\lambda(\lambda,u,v)= -\left.\mu\frac{\partial \lambda}{\partial \mu} \right|_{0},  \qquad
\beta_u(\lambda,u,v)= -\left.\mu\frac{\partial u}{\partial \mu} \right|_{0},  \qquad
  \beta_v(\lambda,u,v)= -\left.\mu\frac{\partial v}{\partial \mu} \right|_{0}, \label{beta}   \\
&&\eta_\psi(\lambda,u,v)= - \beta_{\lambda}(\lambda,u,v)\frac{\partial \ln Z_\psi}{\partial \lambda}
 - \beta_u(\lambda,u,v)\frac{\partial \ln Z_\psi}{\partial u} - \beta_v(\lambda,u,v)\frac{\partial \ln Z_\psi}{\partial v}\,,   \\
&&\eta_m(\lambda,u,v)=
 - \beta_{\lambda}(\lambda,u,v)\frac{\partial \ln Z_m}{\partial \lambda} - \beta_u(\lambda,u,v)\frac{\partial \ln Z_m}{\partial u}
 - \beta_v(\lambda,u,v)\frac{\partial \ln Z_m}{\partial v}\,,\\
&& \gamma(\lambda,u,v) = \eta_m(\lambda,u,v)- \eta_\psi(\lambda,u,v).   \label{gam}
\end{eqnarray}
Here, the subscript ``0'' stands for derivatives at
fixed $\lambda_0$, $u_0$, $v_0$ and $m_0$.
The critical behavior, if present, should be controlled by a stable fixed point (FP)
of the  $\beta$-functions, which is defined as
\begin{eqnarray}
\beta_\lambda(\lambda^*,u^*,v^*)=0, \qquad \beta_u(\lambda^*,u^*,v^*)=0, \qquad \beta_u(\lambda^*,u^*,v^*)=0. \label{eq:fp-def}
\end{eqnarray}
Stability of a given FP can be determined from the
eigenvalues of the stability matrix
\begin{equation}\label{smatrix}
\mathcal{M}=\left(\begin{array}{c c c}
\frac{\partial \beta_\lambda(\lambda,u,v)}{\partial \lambda}&\frac{\partial \beta_\lambda(\lambda,u,v)}{\partial u} & \frac{\partial \beta_\lambda(\lambda,u,v)}{\partial v}\vspace{1mm}\\
\frac{\partial \beta_u(\lambda,u,v)}{\partial \lambda}&\frac{\partial \beta_u(\lambda,u,v)}{\partial u} & \frac{\partial \beta_u(\lambda,u,v)}{\partial v}\vspace{1mm}\\
\frac{\partial \beta_v(\lambda,u,v)}{\partial \lambda}&\frac{\partial \beta_v(\lambda,u,v)}{\partial u} &\frac{\partial \beta_v(\lambda,u,v)}{\partial v}\end{array}\right).
\end{equation}
The FP is stable provided that all the eigenvalues calculated at the FP
(\ref{eq:fp-def}) have negative real parts.
Using coordinates of the stable FP we can calculate the critical exponent. For example,
the correlation length  exponent $\nu$ is given by
 \begin{eqnarray} \label{nufp}
\frac1{\nu} = 1+\gamma(\lambda^*,u^*,v^*).
\end{eqnarray}
The heat capacity exponent is given by the hyperscaling relation, which in two dimensions reads as
\begin{eqnarray} \label{alphafp}
\alpha=2(1-\nu).
\end{eqnarray}
For a marginally irrelevant disorder,
the dependance of the correlation length $\xi:=\re^l$ and the singular part of the heat capacity
$C_{\text{sing}} := \int \rd l F^2 (l) $  on the  reduced temperature $\tau$
can be found from the following flow equations
\begin{eqnarray} \label{flow-exponents-1}
&&\frac{\rd X (l)}{\rd l} = \beta_{X}[\lambda(l),u(l),v(l)], \qquad X=\lambda,u,v, \\
&&\frac{\rd \ln \tau (l) }{\rd l} = -1-\gamma[\lambda(l),u(l),v(l)], \qquad
\frac{\rd \ln F(l) }{\rd l} = \gamma[\lambda(l),u(l),v(l)]. \label{flow-exponents-2}
\end{eqnarray}

\section{One-loop RG flow and the critical behavior} \label{sec:one-loop}

Applying the renormalization procedure described in the previous section
we obtain  the  $\beta$-functions
\begin{align}\label{rg_functions1}
\beta_\lambda(\lambda,u,v)&=\varepsilon\lambda+2(N-2)\lambda^2-4\lambda(u+v),
\\
\beta_{u}(\lambda,u,v)&=\varepsilon u + 4 (N-1) u\lambda-4u(u+v), \label{rg_functions2}
\\
\beta_{v}(\lambda,u,v)&=\delta v + 4 (N-1)v \lambda-4v(u+v), \label{rg_functions3}
\end{align}
and the $\gamma$ function
\begin{eqnarray}
\gamma(\lambda,u,v)= 2(N-1)\lambda-2(u+v)
\end{eqnarray}
to one-loop order in the replica limit $n\to 0$. We find that the flow equations corresponding to
the $\beta$-functions (\ref{rg_functions1})--(\ref{rg_functions3}) for $N>2$ and $\varepsilon>0$
have five distinct FPs:
Gaussian ({\textbf{G}}), Pure ({\textbf{P}}), { \textbf{SR}}, { \textbf{LR}}, and
Mixed~({\textbf{M}}). The coordinates of the FPs  and the eigenvalues $\omega_i$, $i=1,2,3$,
of the stability matrix evaluated at the corresponding FP are summarized
in table~\ref{tab}. Note that in two dimensions, i.e.,   for $d=2$ ($\varepsilon=0$), the first
three  FPs merge and coincide with the    FP  {\textbf{G}}  which describes the critical behavior of $N$
uncoupled clean  Ising models up to logarithmic corrections.
Let us analyze the RG flow for both
the Baxter model ($N=2$) and the
$N$-color Ashkin-Teller model with $N>2$ in three regimes:

\begin{table}[!t]
\centering
\caption{Coordinates of FPs  $\{\lambda^*,u^*,v^*\}$ and the corresponding
stability matrix  eigenvalues $\omega_i$ calculated for the $\beta$-functions
(\ref{rg_functions1})--(\ref{rg_functions3}) for $N>2$.
 \label{tab}}
 \vspace{2ex}
\begin{tabular}{lcccccc}
 \hline
 \hline
  FP & $\lambda^*$ & $u^*$ & $v^*$ & $\omega_1$
  & $\omega_2$ & $\omega_3$ \\ \hline
 {\bf G} & 0 &  0 & 0 & $\delta$ & $\varepsilon$ & $\varepsilon$\\
 {\bf P} &  $-\frac{\varepsilon}{2(N-2)}$ & 0 & 0 &  $\frac{N  (\delta-2\varepsilon) - 2(\delta-\varepsilon) }{ N-2}$ & $-\frac{ N \varepsilon}{ N-2}$ & $-\varepsilon$\\
 {\bf SR} & 0 &  $\frac{\varepsilon}{4}$ & 0 & 0 & $\delta-\varepsilon$ & $-\varepsilon$\\
{\bf LR} & 0 &  0 & $\frac{\delta}{4}$ & $-\delta$ & $-\delta+\varepsilon$ &
 $-\delta+\varepsilon$\\
 {\bf M}& $\frac{\varepsilon-\delta}{2 N }$ &  0 & $\frac{2\varepsilon (N-1)- \delta (N-2)}{4 N}$ &  $-\delta+\varepsilon$ & \multicolumn{2}{c}{$-\frac{\varepsilon}2 \pm   \sqrt {\frac{-4 \delta ^2 (N-2)+4 \delta  (3 N-4) \varepsilon +(8-7 N) \varepsilon
   ^2}{4 N} }$ } \\
\hline \hline
\end{tabular}
\end{table}

(i) \textbf{\textit{ Pure system}} ($u_0=v_0=0$).  For  $N=2$,  the $\beta_\lambda$-function
(\ref{rg_functions1}) vanishes in $d=2$ so that the model has a line of FPs
parameterized by $\lambda$. This is not surprising since
the model in this limit  coincides with the
$O(2)$ Gross-Neveu model or the massive Thirring model.
The $\beta$-function
of the latter model is equal to zero identically leading to
nonuniversal critical exponents \cite{JugShalaev}. This is
consistent with the ``weak universality'' picture given by the exact
solution of the Baxter model~\cite{Baxter}.
Indeed, according to equations~(\ref{nufp}) and (\ref{alphafp}),
the correlation length and the singular part of the heat capacity behave as
\begin{equation}\label{nup-lam}
(N=2): \qquad \xi = \tau^{-1/(1+ 2\lambda_0)}, \qquad
C_{\text{sing}} \sim \tau^{-4\lambda_0},
\end{equation}
where $\lambda_0$ is the initial value of the dimensionless coupling
constant $\lambda$.
Expanding the exponent~(\ref{nup}) in small $J_4$ we find the relation
between parameters of
the continuous and the lattice models: $\lambda_0 \approx 2J_4/(\pi T_{\text c})$.

For $N>2$, the RG flow given by the $\beta_\lambda$-function
(\ref{rg_functions1}) depends on the sign of the initial value of $\lambda_0$:
for $\lambda_0 \leqslant 0$, it flows to zero. Solving the flow
equations~(\ref{flow-exponents-1})--(\ref{flow-exponents-2})  (see the Appendix for details) we find
$\lambda(l)\approx - 1/[2(N-2)l] $ for $l\gg 1$ and  arrive at (see also \cite{Shalaev:1994})
\begin{eqnarray}
(N>2,\ \lambda_0<0): \qquad  \xi \sim \tau^{-1} \left( \ln \tau^{-1}\right)^{(N-1)/(N-2)}, \qquad
C_{\text{sing}} \sim \left( \ln \tau^{-1}\right)^{-N/(N-2)}. \label{Nb2lamneg}
\end{eqnarray}
For $\lambda_0 > 0$, the $\lambda$-flow equation~(\ref{flow-exponents-1}) exhibits a
runway, i.e., the coupling constant $\lambda$
leaves the region in which the perturbative calculations are valid. As a result,
the continuous (within a mean-field approximation) transition
is driven by fluctuations to~\cite{Cardy1,Cardy2}:
\begin{eqnarray}
(N>2,\ \lambda_0>0): \qquad  {\text{first\ order\  phase\  transition}}.
\end{eqnarray}
~The above results may be contrasted with those for the pure 2D Ising model:
$\xi \sim \tau^{-1}$ and $C_{\text{sing}} \sim \ln \tau^{-1}$.

(ii) \textbf{\textit{SR correlated disorder}} ($v_0=0$). We find that
the Gaussian FP (\textbf{G}) is the only stable (marginally) FP in $d=2$.
For $N \geqslant 2$ and $\lambda_0>0$, we rederive (see the Appendix for details) the results
of \cite{Cardy1,Cardy2,Murthy1987}:
\begin{eqnarray}
(N \geqslant 2, \ \lambda_0>0): \qquad  \xi \sim \tau^{-1} \left( \ln \tau^{-1}\right)^{1/2}, \qquad
C_{\text{sing}} \sim  \ln \ln \tau^{-1}, \label{N222}
\end{eqnarray}
coinciding with the results for the 2D Ising model ($N=1$) with SR disorder.
For $N = 2$ and $\lambda_0<0$, we find (see also \cite{DD1})
\begin{eqnarray}
(N =2, \ \lambda_0<0): \qquad  \xi \sim \tau^{-1/(1+2\lambda^*)},
\qquad
C_{\text{sing}} \sim   \tau^{4|\lambda^*|}, \label{N2-lam-neg}
\end{eqnarray}
where $\lambda^* = - |\lambda_0| \re^{-u_0/|\lambda_0|}$.
For $N > 2$ and $\lambda_0<0$, the scaling behavior is the same
as in equations~(\ref{Nb2lamneg}).

\begin{figure}[!t]
\begin{center}
\includegraphics[width=128mm]{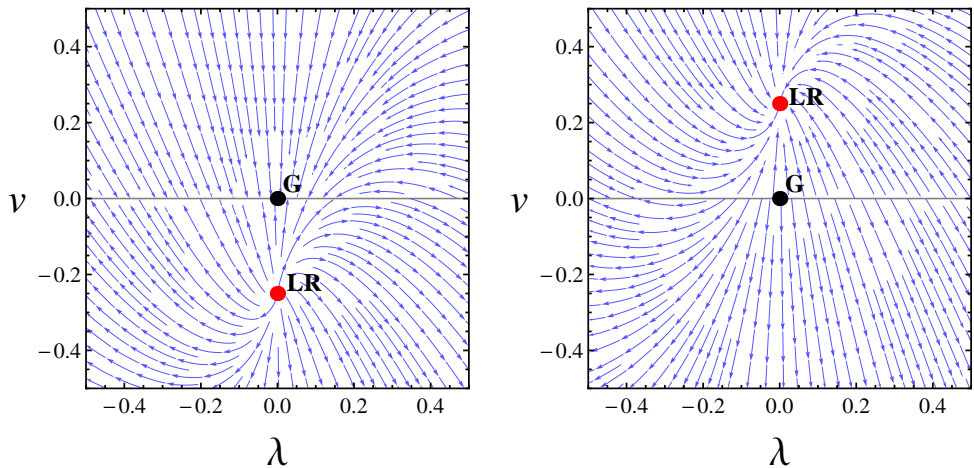}
\end{center}
\vspace{-0.5cm}\caption{ (Color online) The RG flow for the 2D Baxter model ($\varepsilon=0$) with LR correlated disorder in the
plane  $\lambda$, $v$ ($u=0$).
Left-hand panel: $\delta=-1$ ($a=3$). Black dot is the  marginally  stable FP  \textbf{ G}  ($\lambda=u=v=0$).
Red dot is the  FP \textbf{LR} ($\lambda=u=0$, $v=-\frac14$) which is unphysical since $v<0$.
Right-hand panel: $\delta=+1$ ($a=1$). Black dot is the  FP  \textbf{ G} ($\lambda=u=v=0$) which is unstable.
Red dot is the FP  \textbf{ LR} ($\lambda=u=0$, $v=\frac14$) which is stable and physical.     }
  \label{fig:Baxter}
\end{figure}

\begin{figure}[!b]
\begin{center}
\includegraphics[width=128mm]{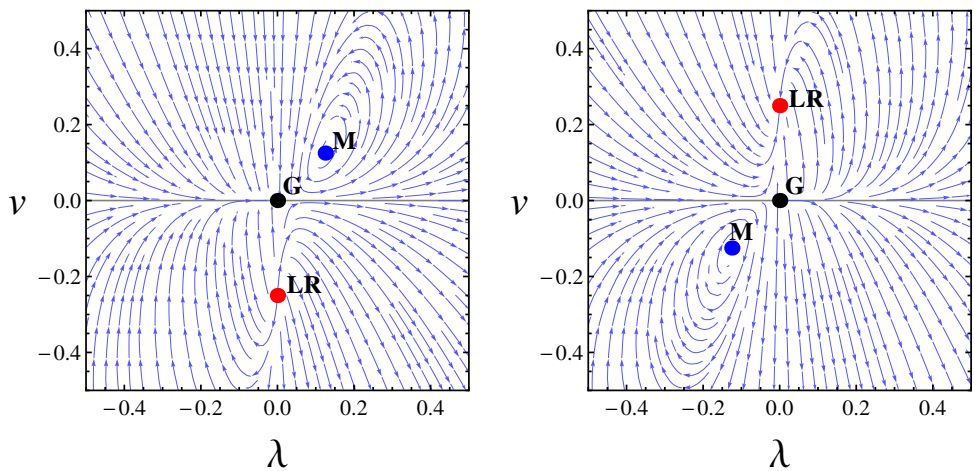}
\end{center}
\vspace{-0.5cm}\caption{(Color online) The RG flow for the 2D 4-color Ashkin-Teller  model ($\varepsilon=0$, $N=4$)
with LR correlated disorder in the plane  $\lambda$, $v$ ($u=0$).
Left-hand panel: $\delta=-1$ ($a=3$). Black dot is the marginally  stable  FP  \textbf{G} ($\lambda=u=v=0$).
Blue dot is the   FP  \textbf{M}  with complex eigenvalues, which is an unstable cycle.
Red dot is the  FP \textbf{LR} ($\lambda=u=0$, $v=-\frac14$) which is unphysical ($v<0$).
Right-hand panel: $\delta=+1$ ($a=1$). Black dot is the  FP  \textbf{G} ($\lambda=u=v=0$) which is unstable.
Blue dot is the   FP  \textbf{M} with imaginary eigenvalues which is an unphysical stable cycle ($v<0$).
Red dot is the  FP \textbf{LR} ($\lambda=u=0$, $v=\frac14$) which is stable and physical.     }
  \label{fig:N-ATM}
\end{figure}

(iii) \textbf{\textit{LR correlated disorder}} ($u_0 \neq 0$, $v_0 \neq 0$). Typical RG flows for
the 2D Baxter  ($N=2$) and the $N$-color Ashkin-Teller  models ($N=4$) are shown
for $\delta=\pm 1$ in figures~\ref{fig:Baxter} and \ref{fig:N-ATM}, respectively.
For $\delta<0$, the LR disorder is irrelevant while  the SR disorder is only
marginally irrelevant. Thus, one can neglect the contribution from the LR disorder
and the scaling behavior is given by
{ equation~(\ref{N222}) for $(N \geqslant 2,\ \lambda_0>0)$,
equation~(\ref{N2-lam-neg}) for $(N =2,\ \lambda_0<0)$ and
equation~(\ref{Nb2lamneg}) for $(N>2,\ \lambda_0<0)$.  }
Note that even if the SR part of disorder is not present in the bare correlator
it will be generated by higher loop order corrections.
For $\delta>0$,  the critical behavior of both models is controlled by the  FP \textbf{LR}:
the models  exhibit the scaling behavior of the 2D Ising model with LR correlated disorder.
For instance, substituting the  FP \textbf{LR} to equation~(\ref{nufp}) we obtain the correlation length exponent
to one-loop:
\begin{equation}
\frac{1}{\nu_{\text{LR}}}=1-\frac{\delta}{2}\,.
\end{equation}
This result was already obtained for the 2D Ising model with LR correlated
disorder in \cite{correlated}
which supports the conjecture that the exact correlation length exponent is $\nu_{\text{LR}}={2}/{a}$.
The corresponding heat capacity exponent is $\alpha_{\text{LR}}=2-a$.

Let us briefly discuss the validity of the extended Harris criterion for the 2D Baxter model with LR correlated
disorder. According to the extended Harris criterion, the LR correlated disorder is relevant provided that
the correlation length exponent of the pure system satisfies the inequality
$a<2/{\nu_{\text{pure}}}$, i.e., $\delta > - 4\lambda_0$.
Although the extended Harris criterion
correctly predicts the relevance of the LR correlated disorder for
 $\delta>0$, as
we  found  above  the  critical  behavior is in fact  modified for any
$\delta<0$. { In  the  last case, FP \textbf{LR} is unstable, and
the asymptotic critical behavior is described by equations~(\ref{N222})
and (\ref{N2-lam-neg}), which corresponds to the 2D Baxter model with SR disorder
rather than to the critical behavior of the pure 2D Baxter model. }
Thus, in the case of the 2D Baxter model,
the extended Harris criterion is
violated by correlated disorder in the same way as the usual Harris criterion is
violated by uncorrelated disorder~\cite{DD1}.

\section{Conclusion}  \label{sec:conclusion}

We have studied the effect of LR correlated disorder on the 2D Baxter
and $N$-color Ashkin-Teller models.
The clean 2D Baxter model exhibits a ``weak universal'' critical behavior with the critical
exponents depending on  microscopic  parameters, while in the clean $N$-color Ashkin-Teller model,
fluctuations drive the system from the second order to the first order phase transition.
Using the mapping to the 2D interacting Dirac fermions in the presence of LR correlated random mass
disorder and dimension regularization with double expansion in $\varepsilon=2-d$ and $\delta=2-a$
we obtain the  RG flow equations to one-loop approximation. Their analysis in  $d=2$ shows that
(i) for $a>2$ ($\delta<0$), the critical behavior is controlled by the Gaussian FP that gives the critical
exponents of the clean 2D Ising model (up to logarithmic corrections);
(ii) for $a<2$ ($\delta>0$), the only stable FP is
 the { LR} FP ($\lambda^*=0,\, u^*=0,\, v^*=\delta/4$). It describes the rounding of the
 weak universality in the  Baxter model and  the  first order phase transition
 in the $N$-color Ashkin-Teller model by correlated disorder.
 This leads to a new emergent critical behavior which is in the same
 universality class as the 2D Ising model with LR correlated  disorder \cite{correlated}.
 For instance, we argue that the exact values of the correlation length and heat capacity
 exponents are $\nu_{\rm LR} = 2/a$ and $\alpha_{\rm LR} = 2-a$, respectively.
 { Since quantum systems can be mapped onto
  classical systems in $d+1$ dimensions, it would be interesting if
  these results could be generalized to the first order phase transitions
  in  1D quantum systems~\cite{quantum-1D1,quantum-1D2}.  }

\section*{Acknowledgements}
It is a great pleasure and a big honor for us to contribute this paper to the festschrift dedicated to
60th Birthday of Yurij Holovatch, who made significant contribution  to understanding scaling properties
of a physical system with quenched disorder, in particular with correlated quenched defects {\cite{Holovatch_rev,correlated,holovatch1,holovatch2,holovatch3,holovatch4,holovatch5,holovatch6,holovatch7,holovatch8}}.

The work of MD was supported in part by ERC grant No. FPTOpt-277998.
MD would like to thank Physics Laboratory of ENS Lyon for hospitality where part of this work was done.
AAF acknowledges support from the French Agence Nationale de la Recherche through
Grants No. ANR-12-BS04-0007 (SemiTopo), No. ANR-13-JS04-0005-01 (ArtiQ),
and No. ANR-14-ACHN-0031 (TopoDyn).

\appendix
\section*{Appendix}

Here, we show how the asymptotic scaling behavior can be derived from the flow equations (\ref{flow-exponents-1}) for $\lambda$ and $u$
\begin{eqnarray} \label{flow-eq-1}
&&\frac{\rd \lambda}{\rd l} = 2(N-2) \lambda^2 -4\lambda u, \qquad
\frac{\rd u}{\rd l} = 4(N-1) \lambda u -4 u^2.
\end{eqnarray}
For $N=2$, we introduce $x =  u/|\lambda|$ that leads to
\begin{eqnarray} \label{flow-eq-10}
&&\frac{\rd \ln x}{\rd l} =  4 \lambda, \qquad
\frac{\rd \ln |\lambda|}{\rd l} = - 4 x |\lambda| .
\end{eqnarray}
For $\lambda_0>0$, we find $x+\ln \lambda ={\text{const}}$. Therefore, the asymptotic behavior
for large $l$  reads
\begin{eqnarray} \label{flow-eq-44}
 u \approx \frac{1}{4l}\,, \qquad  \lambda=\frac1{4l \ln 4l}\,.
\end{eqnarray}
Substituting equation~(\ref{flow-eq-44}) into equations~(\ref{flow-exponents-2}) and noticing that the flow is dominated by
$u(l)$ we obtain equations~(\ref{N222}). For $\lambda_0<0$ we find $x-\ln |\lambda| ={\text{const}}$ and
\begin{eqnarray} \label{flow-eq-444}
&&\frac{\rd |\lambda|}{\rd l} =  -4 \lambda^2(x_0+\ln\lambda/\lambda_0),
\end{eqnarray}
which has a FP solution $\lambda^*=\lambda_0 \re^{-x_0}$, $x^*=u^*=0$.
Substituting this FP into equations~(\ref{nufp}) and (\ref{alphafp})  we obtain the scaling
behavior~(\ref{N2-lam-neg}).

In the case $N>2$ we change variables $x = \pm u/\lambda$ and $y = u^{N-2}/\lambda^{2(N-1)}$, where
``$+$'' and ``$-$'' correspond to $\delta>0$ and $\delta<0$, respectively. This yields
\begin{eqnarray} \label{flow-eq-2}
&&\frac{\rd \ln x}{\rd l} = \pm 2N \frac{x^{(N-2)/N}}{y^{1/N}}\,, \qquad
\frac{\rd \ln y}{\rd l} =  4N \frac{x^{2(N-1)/N}}{y^{1/N}}\,.
\end{eqnarray}
Dividing the first equation by the second one we obtain
\begin{eqnarray} \label{flow-eq-3}
&&\frac{\rd \ln x}{\rd l} = \pm 2N \frac{x^{2(N-1)/N}}{y_0^{1/N}}\re^{\pm 2 (x_0-x)/N}, \qquad
\ln{y/y_0}=\pm 2(x-x_0).
\end{eqnarray}
We define  $x_0 = \pm u_0/\lambda_0$,  $y_0 = u_0^{N-2}/\lambda_0^{2(N-1)}$ and $a_0=\re^{\pm 2x_0}/y_0$.
For $\lambda_0>0$ we find $x^{-2+2/N}\re^{2x/N} \approx 4 l a_0^{1/N}$. Thus, the asymptotic behavior
for large $l$  is given by
\begin{eqnarray} \label{flow-eq-4}
 \  u \approx \frac{1}{4l}\,, \qquad  \lambda=\frac1{2lN \ln 4l}\,.
\end{eqnarray}
Substituting equation~(\ref{flow-eq-4}) into equations~(\ref{flow-exponents-2}) and noticing that the flow is dominated by
$u(l)$ we obtain equations~(\ref{N222}).
For $\lambda_0<0$, we find $x^{-1+2/N} \approx 2 (N-2) l a_0^{1/N}$ which leads to the following
asymptotic behavior
\begin{eqnarray} \label{flow-eq-5}
 \  u \approx \frac{a_0^{-1/(N-2)}}{[2(N-2)l]^{2(N-1)/(N-2)}}\,, \qquad  \lambda=-\frac1{2(N-2)l}\,.
\end{eqnarray}
Substituting equation~(\ref{flow-eq-5}) into equations~(\ref{flow-exponents-2}) and noticing that the flow is dominated by
$\lambda(l)$ we obtain the same scaling behavior as in equations~(\ref{Nb2lamneg}).

\newpage
\ukrainianpart

\title{Універсальна критична поведінка 2D $N$-кольорової моделі Ашкіна-Телера, що з'являється у присутності довгосяжно-скорельованого безладу}
\author{М. Дудка\refaddr{label1}, А.А. Федоренко\refaddr{label2}}
\addresses{
\addr{label1} Інститут фізики конденсованих систем НАН України,
вул. Свєнціцького, 1, 79011 Львів, Україна
\addr{label2} Лабораторія фізики, Вища нормальна школа м. Ліон, Університет Клода Бернара, ЦНРС,\\  F-69342 Ліон, Франція
}

\makeukrtitle

\begin{abstract}
\tolerance=3000%
Ми вивчаємо критичну поведінку 2D $N$-кольорової моделі Ашкіна-Телера у присутності безладу типу випадкових зв'язків, кореляції якого
спадають з відстанню $r$ за степеневим законом $r^{-a}$. Ми розглядаємо випадок, коли спіни різних кольорів, що сидять на тому ж самому вузлі, зв'язані одним зв'язком, і переводимо цю задачу на 2D систему $N/2$ ароматів взаємодіючих ферміонів Дірака у присутності скорельованого безладу. Використовуючи ренорм-групу, ми показуємо, що для випадку $N=2$ ``слабка універсальність'' при неперервному переході стає універсальною з новими критичними показниками. Для $N>2$ фазовий перехід першого роду перетворюється скорельованим безладом в неперервний.
\keywords фазові переходи, скорельований безлад, двовимірні моделі, ферміони Дірака, ренорм-група

\end{abstract}

\end{document}